\documentclass[twocolumn]{article}
\usepackage[utf8]{inputenc}
\usepackage[T1]{fontenc}
\usepackage{titling}
\usepackage{graphicx}
\usepackage{amsmath}
\usepackage{float}
\usepackage[linesnumbered,ruled]{algorithm2e}
\usepackage{hyperref}
\hypersetup{
  colorlinks=true,
  linkcolor=blue,
  filecolor=magenta,   
  urlcolor=cyan,
}

\title{The Design and Implementation of\\
Modern Online Programming Competitions}

\author{Benjamin Spector \\ \href{mailto:benjaminfspector@gmail.com}{benjaminfspector@gmail.com}
    \and Michael Truell \\ \href{mailto:mntruell@gmail.com}{mntruell@gmail.com}}
\date{October 2017}

\begin{document}

\maketitle

\begin{abstract}
This paper presents a framework for the implementation of online programming competitions, including a set of principles for the design of the multiplayer game and a practical framework for the construction of the competition environment. The paper presents a successful example competition, the 2016-17 Halite challenge, and briefly mentions a second competition, the Halite II challenge, which launched in October 2017.
\end{abstract}

\section{Introduction}

Online programming competitions are a popular way for programmers to learn, collaborate, and relax. For schools and universities, they are a useful immersive tool to motivate and educate students. For companies, they are a route to gain publicity and recruit new programmers as they signal a fun, programmer-centered culture. \cite{airhockey} In a research context, they serve as de facto benchmarks of algorithm performance \cite{mario, gym}. They are popular items in industry, as they attract new hires by signaling a fun, programmer-centered culture \cite{airhockey}.

A programming competition consists of the game itself, which contestants write programs to play, and the surrounding competition infrastructure. 
While much work has been done to formalize the design and implementation of robust, human-playable games, as in Crawford's book which first appeared in the 1980's \cite{CrawfordGameDesign}, this paper serves to formalize a set of requirements and characteristics of a good game-based programming competition. This framework is the Effective Programming Competition Framework (EPCoF).

EPCoF is based on the authors' experiences with a successful, turn-based strategy game called Halite. The design and implementation of Halite served to hone the definition of EPCoF and also to demonstrate its utility. Halite launched publicly in the fall of 2016 \cite{HaliteV1}, had more than 1600 finishers, and received numerous contributions from an engaged open source development community. A second competition, called Halite II and also developed consistently with EPCoF, launched in October 2017 \cite{HaliteV2}.

\section{Effective Programming Competition Framework}

This section presents a framework for the design of a programming competition. This includes both the principles underlying a good programming game, as well as the principles for developing the surrounding programming competition environment. 

\subsection{Principles of Programming Game Design}

The EPCoF principles for the design of the programming game are as follows:

\begin{enumerate}
  \item The game should be simple, intuitive, and easy to start to play. Few programmers will spend hours learning complex rules and setting up an environment; the game and its accompanying environment need to be easy to understand and set up in under ten minutes, regardless of platform or user background.
  \item The game should be visually appealing in order to draw in players and minimize the bounce rate from the website. Making games easy to view and understand also makes it more straightforward for competitors to view games and better their bots, which improves the user experience.
  \item The game design should allow a runtime environment that is computationally inexpensive. Because there can be hundreds of thousands to millions of games, games should be quick to run even on lightweight servers.
  \item The game should be as difficult as possible to perfectly solve. Little can kill a competition more thoroughly than a player discovering a simple dominant strategy. 
  \item The game should always have a path forward for users to improve their bot so that they stay engaged. Whether a user is just beginning to set up a local game environment or vying for the leaderboard's diamond slots, they should feel there is always something to be made better about their bot, because users who feel they have reached a dead end are much less likely to continue to participate. Concretely, this means that the game’s strategy must be many-faceted, so that once a player has optimized an aspect of their bot as best as they can, they can come back to it later and turn to the next one.
\end{enumerate}

\subsection{Principles of Programming Competition Construction}

The EPCoF principles for the design of the actual competition environment are as follows:

\begin{enumerate}
  \item The competition environment should be beginner-friendly. This should be in conjunction with the game design; just as the game must be simple to learn, the platform in which the game is programmed, debugged, and run must be simple to use.
  \item The competition platform must be secure to minimize the possibility of abuse. This is challenging because implementations will typically allow players to upload executable logic into a shared environment.
  \item The competition platform must easily scale with the number of contestants.
  \item The competition should provide real-time standings of contestants in order to show players their progress and encourage their continued interest.
  \item The competition platform should encourage collaboration and community. Collaboration can take many forms, but due thought needs to be given to forums, social media, and the appropriate use of shared code repositories. 
  \item The competition must allow users to easily develop and test their bots. Speed of development, turn-around time in debugging, and flexibility of programming language use are all important objectives.
\end{enumerate}

\section{The 2016-17 Halite Competition}

The Halite game started with an initial focus on the game challenge. Through various iterations over the course of about 12 months and with successively larger pilot usage, the game itself converged and more attention became focused on having a competition environment that could support thousands of users. Early on, the authors decided to develop Halite in an open source manner hosted on GitHub, a valuable decision that allowed the user community to contribute enormously to the success of the game -- particularly in increasing support for the game to about a dozen programming languages. Halite served both as a basis for refining the EPCoF framework, but also became a good case study of its utility. 

\subsection{The Halite Game}

\subsubsection{Rules}

Halite is a turn-based strategy game in which between two and six players participate. Halite is played on a wrap-around (toroidal) rectangular grid, and each player has the objective of eliminating all other players from the grid. Each player starts with a single piece owned on the grid, and expands their territory by capturing unowned or enemy-owned sites. With every turn, each player is sent the complete Halite game state and is expected to return their set of moves for their pieces. The game engine then executes all moves simultaneously. Halite maps range from 20x20 to 50x50 and usually take several hundred turns to finish.

\begin{figure}[H]
  \centering
  \includegraphics[width=0.8\linewidth]{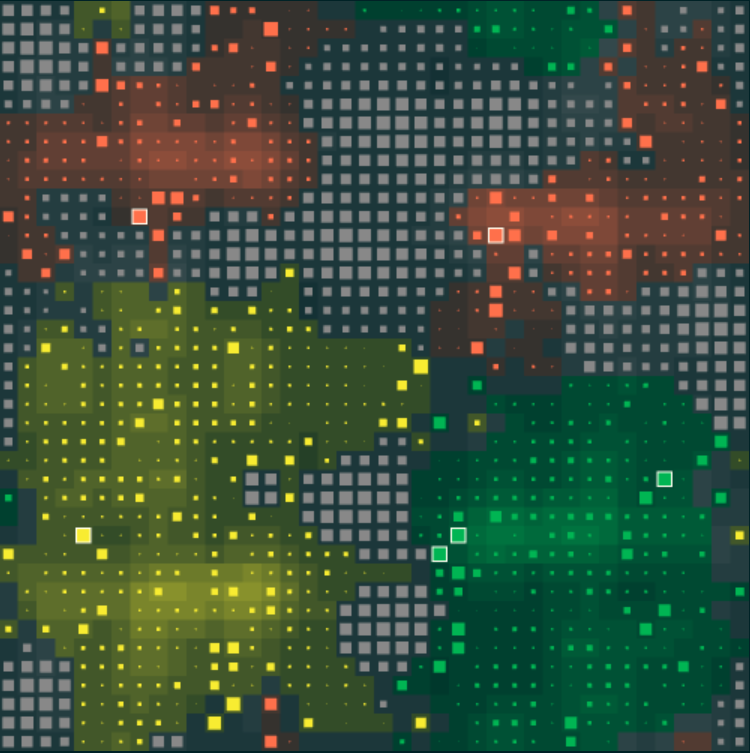}
  \caption{A snapshot of the game through its visualizer. In this game, the orange player has appreciably less territory or strength than its opponents, but considerably greater production, represented by the brighter regions of the background map.}
\end{figure}

During a turn, each piece may be moved in one of the four cardinal directions or remain still, and piece additionally has an integer strength associated with it (this strength is capped at 255; any excess strength is wasted). When a piece is ordered to be still, its strength is increased, and when a piece moves, it leaves behind a piece with the same owner and a strength of zero. When two or more pieces from the same player try to occupy the same site, the resultant piece has a strength equal to the sum of their strengths. However, when pieces with different owners move onto the same site or cardinally adjacent sites, the pieces must fight, and each piece loses strength equal to the strength of its opponent. When a player's piece moves onto an unowned site, it will lose an amount of strength equal to the strength of the unowned piece, and take the site if it remains alive. Additionally, each piece does all of its attacks before it takes any damage, meaning it can damage multiple pieces simultaneously. When a piece loses all of its strength, it dies and is removed from the grid.

The last element of Halite is its maps: every site on the map has associated with it a starting strength and a value called production, which defines how fast owned pieces grow when they are are ordered to remain still. Not all sites on a Halite map are created equal -- a site with a small strength and high production is very valuable, as it costs little to obtain but rapidly produces strength, whereas a site with high strength and low or no production may not ever be worth taking. Halite maps are randomly generated using a custom algorithm designed to create interesting ones, and no two games are ever played on the same map.

\subsubsection{Analysis}

Fundamentally, playing Halite involves optimizing and balancing four major objectives:

\begin{itemize}
  \item The bot must effectively expand its territory. This requires choosing the territory towards which the bot will move while balancing the gaining of production as well as the preservation of bot strength. This turns out to be a surprisingly interesting optimization problem.
  \item The bot must strategically take territory from other bots. A bot must choose how much strength to devote to attacking an enemy. If a bot devotes too little, it will lose on those fronts and quickly be overrun, but if it devotes too much, it will be less effective at expanding elsewhere and will also lose.
  \item A bot must be tactically effective in combat, maximizing damage inflicted on an opponent, as well as minimizing damage taken from said opponent. Additionally, a bot may want to occasionally sacrifice some additional strength in order to break through a front, because damage done in this way becomes much harder to contain.
  \item A bot should be efficient in building up its strength. Because pieces only gain strength when stationary, this entails ordering them to be still as often as possible. However, due to the 255 strength cap, the flow of pieces to the borders of the bot can often become clogged if there is too little movement. This may lead to great inefficiencies within the bot.
\end{itemize}

In addition to optimizing for those core aspects of Halite, bots ought to deal with other characteristics and emergent behaviors of Halite to rise in the rankings. There are too many to list here, so some selected features are summarized below:

\begin{figure}[H]
  \centering
  \includegraphics[width=0.8\linewidth]{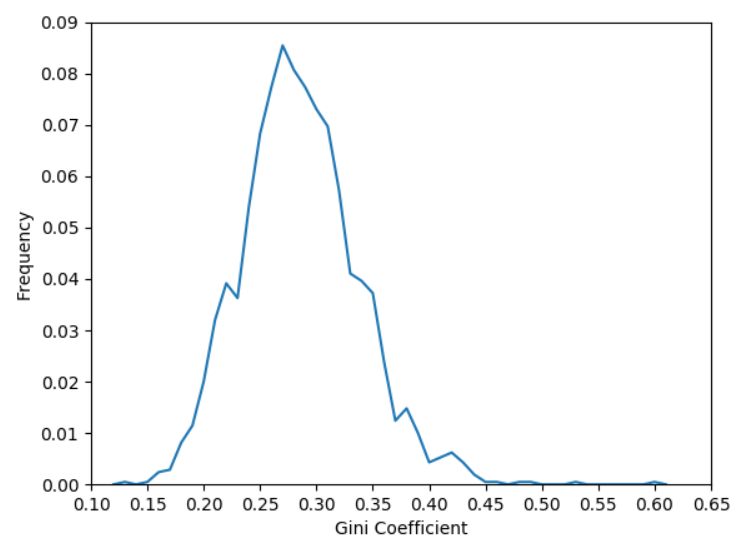}
  \caption{A plot of the frequency of Gini coefficients (rounded to the nearest 0.01) in a random sample of 2095 Halite games.}
\end{figure}

Halite's map generator is itself an element of the game and produces a variety of interesting maps, challenging users to build versatile and adaptive bots. One way of characterizing a map is by the distribution of production throughout the map. Some maps have their production fairly evenly distributed throughout the map, whereas others have virtually all of it concentrated in specific regions of the map. An easy way to numerically characterize this is to borrow the Gini coefficient from economics, where a coefficient of 0 corresponds to a perfectly even distribution and a coefficient of 1 represents a single site having all of the production of the map. Figure 2 shows a coefficient distribution centered at a approximately 0.27-0.28, but with considerable variation. This indicates that a Halite bot should be optimized for that class of maps, but still have the versatility to deal with other types as well.

\begin{figure}[H]
  \centering
  \includegraphics[width=0.8\linewidth]{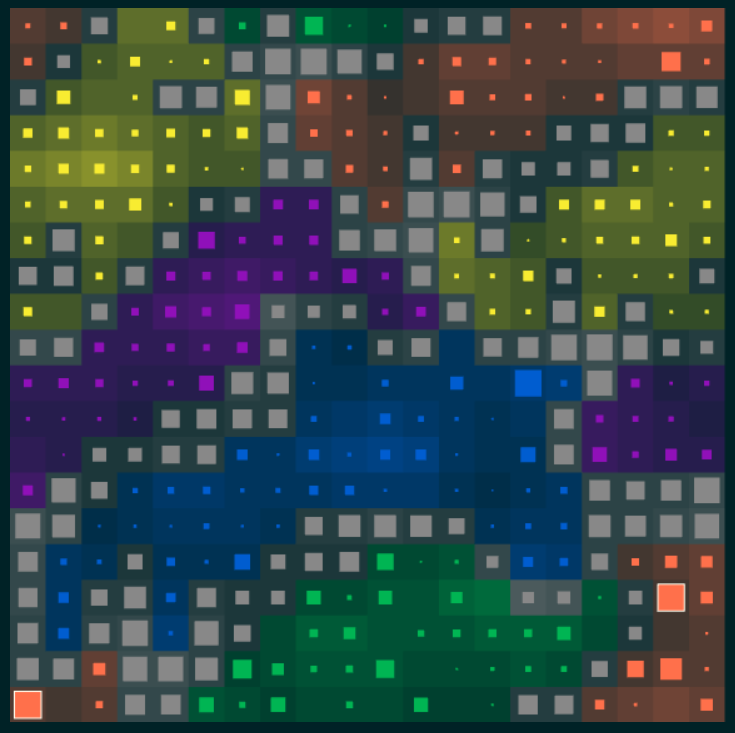}
  \caption{A snapshot of a game in which all players are obeying the non-aggression pact.}
\end{figure}

The non-aggression pact (NAP) is an emergent phenomenon that was discovered in the final weeks of the competition and exploded during the last few days. It states that in a multiplayer game, bots should always let an opponent be the first to attack them. When playing only with players not following the NAP, this has essentially no effect because very little happens in Halite during any single turn. However, when two or more NAP-enabled bots participate in the same match, each is helped because they make peace with each other and do not fight until they are among the last bots remaining (as may be seen in figure 3), meaning they can direct more strength at their other opponents. This turns out to be a highly significant effect, and was responsible for a great deal of volatility in the rankings during the last few days of the competition.

\subsubsection{EPCoF Compliance}

Halite satisfies the EPCoF requirements outlined above to a great degree.

First, it is a remarkably uncomplicated game. Players just send one simple action to every piece, every turn, and the complexity of the game arises from the interactions of the rules rather than the rules themselves.

Second, Halite follows good principles for visual design. Because the game is played on a grid, the games are easy to visualize, and make use of bright primary colors to draw in players. Additionally, the continuous nature of players' territories makes it easy to see the tempo of the game, allowing details to be ignored and the big picture to be seen.

Third, Halite is indeed very fast to compute. The game engine computes the results of each turn in O(n), where n is the number of tiles on the map, and the constant factor is very small, too. Because there were never problems due to the speed of the un-optimized C++ (for it was still very fast), the code was not further improved, though a user, arjunvis, rewrote and optimized the environment in MATLAB in order to create his bot and reported a simulation speed of 20,000 turns per second.

Fourth, Halite is essentially unsolvable. Halite maps introduce considerable variety to the game, and the branching factor for Halite may be on the order of $10^{1743}$ for the largest maps, making any sort of exhaustive tree search unattainable. This is also made more difficult by the fact that it is not straightforward to reliably evaluate the value of a Halite board for a player. Heuristics, such as the global sum of each player’s territory, strength, and production, are relevant but not decisive, because a great number of other factors go into correctly evaluating a board. It may be possible to use evolutionary strategies to generate move-sets as described in Justesen et. al. \cite{justesen2016}, but to our knowledge no competitors used this strategy because direct policy is much simpler to execute and very effective at playing Halite -- an aspect that made Halite more fun to play. 

Last, Halite encourages iterative bot improvement. Because each of the four major objectives of a bot can be achieved in many ways, optimized independently or together, and each objective consists of dozens of sub-problems, there is always a path in sight to a better Halite bot.

For the record, considerable experience was needed to get the game design right. The game designed evolved several times while trying to meet this specific requirement.

\subsection{The Halite Competition Construction}

This section discusses the technical and user-oriented aspects of the Halite competition and its compliance with EPCoF.

Users built their bots using the programming language of their choice locally using the Halite game environment software. Once users were happy with their bots, they submitted their source to the competition website. The game servers would compile their source, run the resulting binary against other bots, and rank it on a global leaderboard. Replays of all server-side games were made publicly available through the site, and they remain available. Throughout the competition, users could make improvements to their bots and resubmit at their leisure. The winner of the competition was the user who led the leaderboard at the end of the three months.

\subsubsection{Bot Development}

Halite satisfied the first requirement of EPCoF, namely that users can effectively build bots for the competition. 

Users were provided with a local game binary that would execute Halite game logic and serve as the intermediary between competing bots. Bot programs communicated with this game environment using pipes, allowing the environment to interface with any programming language. Users were provided with small libraries for Java, C++, and Python to get them started with building a bot. These libraries, called "starter packages," consisted of an implementation of the environment's pipe networking protocol, simple game object representations, and a dummy Halite bot that moved all of its game pieces randomly (see the above Rules section). The game environment exposed a flexible command line interface, which allowed users to build their own tools for automating parts of the development process. For example, one user created a script to rank his various local bots according to the Trueskill algorithm \cite{trueskillpaper}, in the manner of Halite's central leaderboard.

\subsubsection{Ranking}

Throughout the entirety of the Halite competition, users had a real-time centralized, public ranking showing their bots performance. Users submitted their local bot source code to our competition website, which would then compile their bots and run them against other the bots of other users. The competition platform would publicly display replays of these games and rank entrants according to the Trueskill ranking algorithm \cite{trueskillpaper}. Users were permitted to submit updates to their bots at their leisure.

\begin{figure}[H]
  \centering
  \includegraphics[width=0.95\linewidth]{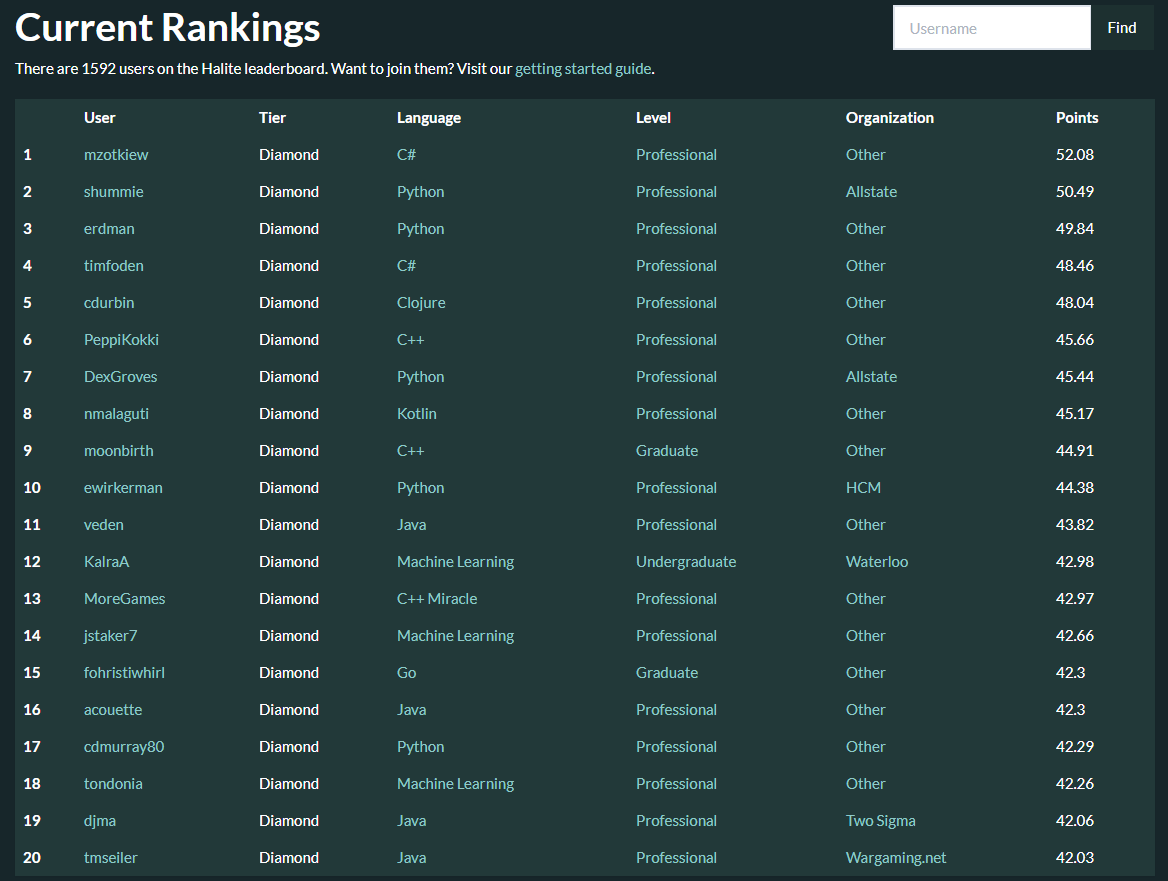}
  \caption{A snapshot of the final rankings of top Halite participants. The columns can be clicked to sort. Note that Halite II has devoted considerable focus to making it more natural for users to list an organizational affiliation.}
\end{figure}

We believe that this leaderboard system was one of the most important features of the Halite competition. It allowed users to gauge their progress in building a bot, pushed users to build more ambitious bots, and served to disseminate bot algorithm ideas (via the associated game visualizations) throughout the Halite user base.

\subsubsection{Scaling}

The competition platform was built to scale well as the user base grew. Namely, as more bots were submitted, the ranking platform had to expand. Thus, ranking servers were organized in a one-to-many server architecture. An arbitrarily scalable field of worker servers performed the actual running of ranking games and compilation of bot source. A single manager server coordinated the jobs of these workers -- telling them which bots to compile or which games to run -- and served as the gatekeeper to a single consistent repository of bot rankings and competition source. 

The ranking servers were hosted on AWS EC2. They interfaced with the AWS EC2 API via the boto library to start up new servers every time a constant number (set to 50) of new users joined the service.

\subsubsection{Security}

The Halite competition platform was built to be secure.

The execution and sandboxing of users' bot code on Halite servers was of primary concern. To accomplish this, Docker containers were used to isolate user-submitted bots from each other and the game environment, limit their CPU and memory resources, and prevent them from accessing the network. Docker was also used to sandbox compilation tasks. While initial designs considered using VMs and chroot jails as the sandboxes, these two ideas were eventually discounted for their latency and insecurity, respectively.

To the best of our knowledge, we experienced no issues with security. Players expressed little interest in winning subversively, favoring instead to develop their bots and play legitimately.

\subsubsection{Community}

In order to build a community, the website hosted a Discourse forum. The community built around this forum was vibrant, resulting in thousands of posts and hundreds of thousands of page-views. Additionally, the community self-started an active Discord channel, in which players chatted and collaborated. Several open voice chats between players were also arranged.

Last, we ran a Halite hackathon for high school students at the Dalton School in New York City. This helped younger students get into the game and start participating, and it highlighted the educational aspect of the competition. One high school student team did very well. The educational value has also been demonstrated by others using Halite; for example, the Federal University of Paraná in Brazil recently decided to use Halite to challenge some students with the caveat that they will need to demonstrate the use of certain interfaces and algorithms. \cite{BrazilHalite}

The end result of this was a community that was very active and cared deeply about Halite. Users submitted their own starter packages, revised game documentation, provided additional tutorials, and did extra analysis of the website and the game. All of this contributed immensely to the end user experience.



\subsubsection{Beginner-Friendliness}

Halite had a number of features that made it very beginner-friendly.

First, Halite single sign-on (SSO) was integrated with GitHub. This meant that any potential contestant could sign up in seconds with a one-click option, encouraging participation of players new to Halite.

Second, the documentation for the game was clear, concise, and informative. There were tutorials, in both written and visual form, which took users through the process of setting up the environment, building a basic bot, and submitting it.

Third, the centralized leaderboard helped beginners track their progress. Tiers allowed beginners to make goals (there was much discussion of how to become "diamond-tier"), encouraging them by showing both how far a bot can go, as well as the path for getting there.

Last, by providing starter packages, we simplified the process of writing a bot for the user by eliminating boilerplate code and additionally provided users with a stable bot they could immediately submit to get onto the leaderboard without even touching a single line of code. The official Python, Java, and C++ provided from the start of the competition accounted for >80\% of all eventual users, and Python by itself was used by over half of all users. Nonetheless, users contributed numerous packages for other languages, resulting in packages for 11 additional languages by the end of the competition.


\section{Conclusion}

This paper details the construction and administration the Two Sigma and Cornell Tech Halite Challenge, a successful online programming competition. We've presented our framework for building online programming competitions, both from a general perspective as well as a specific case study. In the future, we hope that others will use, modify, and revise these guidelines in the process of building similar competitions.

\section*{Acknowledgements}

We would like to thank Emily Malloy, Jacques Clapauch, Matthew Adereth, Herbert Wang, and Arnaud Sahuguet for their invaluable help in running the competition. We’d also like to thank Trammell Hudson and Steve Heller for their mentorship, Alfred Spector for his endless advice, and Two Sigma Investments in its entirety for sponsoring, participating in, and wholeheartedly embracing the concept behind Halite. We thank the Ants AI Challenge, which inspired both the Halite game and competition. We also appreciate Nikki Ho-Shing's editorial comments. Last, we thank our competitors for building a fun, supportive, and intelligent community. Special thanks goes to our community contributors, most notably Brian Haskin, Travis Erdman, and Nick Malaguti. 

We encourage the reader to participate in Halite II starting in October 2017 \cite{HaliteV2}, as developed by David Li, presently at Cornell, and Jaques Clapauch, Harikrishna Menon, and Julia Kastner, employed by or consulting for Two Sigma. 

\section*{About the Authors}

Benjamin Spector and Michael Truell are presently at the Horace Mann School in Bronx, NY.  Much of this work was done at Horace Mann, but the full-fledged production of Halite was done at Two Sigma and benefited greatly from its support.

\nocite{*}
\bibliographystyle{ieeetr}
{\small \bibliography{ref}}

\end{document}